\documentclass[preprint,nofootinbib,floatfix,a4paper,prd,superscriptaddress]{revtex4-1}   	
\usepackage{geometry}                		
\usepackage[colorlinks,citecolor=blue]{hyperref}
\usepackage{amsmath}
\usepackage{mathrsfs}
\usepackage{bbm}
\usepackage{amsfonts}
\usepackage{amssymb}
\usepackage{latexsym}
\usepackage{graphicx}
\usepackage[english]{babel}
\usepackage{multirow}
\usepackage{float}
\usepackage{url}
\usepackage{slashed}
\usepackage{ulem}

\usepackage{orcidlink}

\newcommand{\be}{\begin{equation}}
\newcommand{\ee}{\end{equation}}
\newcommand{\ba}{\begin{align*}}
\newcommand{\ea}{\end{align*}}
\newcommand{\bea}{\begin{eqnarray}}
\newcommand{\eea}{\end{eqnarray}}

\usepackage[utf8]{inputenc}
\usepackage[T1]{fontenc}

\newcommand{\PU}{\affiliation{\small Phenikaa Institute for Advanced Study, Phenikaa University, Nguyen Trac, Duong Noi, Hanoi 100000, Vietnam}}

\newcommand{\NCTS}{\affiliation{\small Physics Division, National Center for Theoretical Sciences, National Taiwan University, Taipei 106319, Taiwan}}

\newcommand{\NTNU}{\affiliation{\small Department of Physics, National Taiwan Normal University, Taipei 116, Taiwan}}

\begin{document}

\title{Probing the Axion--Photon--Dark Photon Interaction \\ at Future $e^+e^-$ Colliders}

\author{Chuan-Ren Chen}
\email{crchen@ntnu.edu.tw} \NTNU

\author{Yuan-Feng Hsieh}
\email{61041026S@ntnu.edu.tw} \NTNU

\author{Van Que Tran \orcidlink{0000-0003-4643-4050}}
\email{vqtran@phys.ncts.ntu.edu.tw} \NCTS \PU


\begin{abstract}
We study the interaction between photons, dark photons, and axions at future lepton colliders, focusing on single-photon events with missing energy as the experimental signature. We find that future facilities such as the ILC, CEPC, and FCC-ee will be sensitive to the axion--photon--dark photon coupling down to the order of $10^{-4}\, \mathrm{GeV}^{-1}$ for dark photon masses around $\mathcal{O}(10~\mathrm{GeV})$, assuming that the axion is extremely light and escapes detection. We further show that longitudinal beam polarization at the ILC can enhance the signal significance by a factor of four, providing the 
strongest projected reach in the model parameter space.
Existing constraints from LEP II are analyzed for comparison. Furthermore, the mass of dark photon can be determined by measuring the sharp drop-off in the distribution of the recoil mass.
\end{abstract}

\maketitle

\section{Introduction}

The exploration of light and weakly coupled particles beyond the Standard Model (SM) is interesting on its own in the search for new physics. Among the most theoretically motivated candidates are axions, originally proposed to solve the strong CP problem (see, e.g.~\cite{Choi:2020rgn} for a review), and dark photons, which arise as gauge bosons of hidden $U(1)$ symmetries appearing in many extensions of the SM, including string-theoretic and grand unification models (see, e.g.~\cite{Fabbrichesi:2020wbt} for a review).

Both axions and dark photons are typically light and feebly interacting with the visible sector through higher-dimensional operators or a kinetic mixing term. Of particular interest is the possibility of an interaction vertex involving an axion, a photon, and a dark photon, which can arise naturally in UV completions involving heavy fermions charged under both SM and dark gauge groups~\cite{Kaneta:2016wvf,Zuowei:2025}.
This interaction gives rise to a rich phenomenology not only in astrophysical and cosmological contexts~\cite{Kaneta:2017wfh, Choi:2018mvk, Kalashev:2018bra, Choi:2019jwx, Hook:2019hdk, Arias:2020tzl, Hook:2021ous, Domcke:2021yuz, Gutierrez:2021gol, Carenza:2023qxh, Hook:2023smg, Hong:2023fcy, Broadberry:2024yjw}, but also in collider and reactor experiments, where it leads to distinctive signatures beyond those induced by kinetic mixing alone~\cite{Biswas:2019lcp, Deniverville:2020rbv, Lane:2023eno, Jodlowski:2023yne}. Moreover, the interaction of the axion--photon--dark photon has also been studied in various other contexts~\cite{Chen:2024jbr, Ding:2025eqq}.

In this work, we explore the phenomenology of the axion--photon--dark photon interaction at LEP II and future lepton colliders. In particular, we study processes $e^+ e^- \to \gamma'$ and $e^+ e^- \to \gamma' a$ followed by decay $\gamma' \to \gamma a$, where $\gamma'$ and $a$ represent dark photon and axion, respectively. These processes yield the same collider signature consisting of a single photon plus missing energy, as the axion escapes detection. The single-photon signature is especially clean at lepton colliders because of the relatively low and well-understood backgrounds. 
We analyze existing data from LEP II to place constraints on the axion--dark photon--photon coupling and then project the sensitivity of future machines such as the International Linear Collider (ILC)~\cite{ILC:2013jhg, LCCPhysicsWorkingGroup:2019fvj} , the Future Circular Collider (FCC-ee)~\cite{FCC:2018evy}, and the Circular Electron--Positron Collider (CEPC)~\cite{CEPCStudyGroup:2023quu}. In addition, we demonstrate that the recoil mass spectrum in these events contains a distinctive maximum feature that can be used to reconstruct the mass of dark photon.
We also demonstrate the enhancement of the signal significance using the longitudinal polarized beams at the ILC.

This paper is organized as follows. In Section \ref{sec:themodel}, we present the theoretical framework; in Section \ref{sec:signature}, we analyze the signal at LEP and future colliders; in Section \ref{sec:pol} we analyze the signal at ILC with polarized beams; finally, we conclude our findings in Section \ref{sec:conclusion}.

\section{\label{sec:themodel} The Model}

We consider a dark sector involving a light pseudo-scalar $a$ and a dark photon $\gamma'$ that originate from a broken $\text{U(1)}_D$ global symmetry and a broken $\text{U(1)}_X$ gauge symmetry~\cite{Kaneta:2016wvf, Zuowei:2025}, respectively. The relevant interactions can be described by an effective Lagrangian that includes both kinetic mixing and a dimension-five operator involving the axion, dark photon, and SM photon:
\begin{equation}
\label{eq:Lag}
\mathcal{L} \supset -\frac{1}{4} F_{\mu\nu} F^{\mu\nu} - \frac{1}{4} F'_{\mu\nu} F'^{\mu\nu} - \frac{\varepsilon}{2\, {c_W} } F_{\mu\nu} F'^{\mu\nu} + \frac{ g_{a\gamma'\gamma} }{2\, {c_W}}a F_{\mu\nu} \tilde{F}'^{\mu\nu}, 
\end{equation}
where $c_W$ stands for cosine of the weak mixing angle, and $F_{\mu\nu}$ and $F'_{\mu\nu}$ are the field strength tensors of the hypercharge and the dark photon, respectively, and $\tilde{F}'^{\mu\nu} \equiv \frac{1}{2} \epsilon^{\mu\nu\rho\sigma} F'_{\rho\sigma}$ is the dual field strength. The coupling $g_{a\gamma'\gamma}$ has a dimension of $1/\text{mass}$, and its magnitude depends on the UV completion, such as heavy fermion loops or anomaly-induced interactions.

Due to kinetic mixing, the dark photon can decay into SM particles. 
The partial decay width of the dark photon into a pair of charged leptons is given by:
\begin{equation}
\label{eq:Aptoll}
\Gamma(\gamma'\to \ell^+\ell^-) =
\frac{\epsilon^2 e^2}{12 \pi } m_{\gamma'}\left(1+ \frac{ 2 m_\ell^2}{m_{\gamma'}^2} \right) \sqrt{1- \frac{4 m_\ell^2}{m_{\gamma'}^2}} ,
\end{equation}
where $e$ is the electromagnetic coupling, and $m_{\gamma'}$ and $m_\ell$ are masses of the dark photon and lepton $\ell$, respectively.

For hadrons, partial decay widths can be related to the muon channel via:
\begin{equation}
\label{eq:Aptohad}
\Gamma (\gamma' \to \text{hadrons}) = \Gamma( \gamma' \to \mu^+\mu^-) \times R(m_{\gamma'}^2),
\end{equation}
where the function
$$
R(m^2_{\gamma'}) \equiv \frac{\sigma(e^+e^-\to \text{hadrons})}{\sigma(e^+e^-\to \mu^+ \mu^-)},
$$
encodes the effects of hadronic resonances and the mixing of the dark photon with QCD vector mesons. The values of $R(m^2_{\gamma'})$ can be taken from measurements~\cite{ParticleDataGroup:2024cfk}.

As $\gamma'$ is heavy enough, it could decay into a SM $W$-boson pair. The partial decay width is 
\begin{equation}
\Gamma(\gamma'\to W^+W^-)=\frac{\epsilon^2 e^2}{192\pi}m_{\gamma'}\left(\frac{m_{\gamma'}}{m_W}\right)^4\left(1-4\frac{m_W^2}{m_{\gamma'}^2}\right)^{3/2}\lambda(x),
\end{equation}
where $\lambda(x)=1+20x^2+12x^4$ with $x=m_W/m_{\gamma'}$. 

In addition, the dimension-five operator allows the decay $\gamma' \to \gamma a$ and $\gamma' \to Z a$ when kinematically allowed, with partial widths given by:
\bea
\Gamma(\gamma' \to a \gamma) &=& \frac{g_{a\gamma'\gamma}^2}{96 \pi} m_{\gamma'}^3 \left( 1 - \frac{m_a^2}{m_{\gamma'}^2} \right)^3, \\
\Gamma(\gamma' \to a Z) &=& \frac{t_W^2 g_{a\gamma'\gamma}^2}{96 \pi} m_{\gamma'}^3 \left\{ \left[ 1 - \left(\frac{m_Z - m_a}{m_{\gamma'} }\right)^2 \right] \left[ 1 - \left(\frac{m_Z + m_a}{m_{\gamma'} }\right)^2 \right] \right\}^{3/2},
\eea
where $t_W$ is the tangent of the weak mixing angle and $m_a$ is the mass of the axion.
In the kinematically allowed region and for a light axion, the decay 
$\gamma' \to a\gamma$ dominates over $\gamma' \to aZ$ because the latter 
is suppressed by $t_W^2$ and the phase space factor. Furthermore, when 
$m_{\gamma'}\, g_{a\gamma'\gamma} \gtrsim \epsilon$, the partial width 
$\Gamma(\gamma' \to a\gamma)$ exceeds the decay widths into SM leptons 
and hadrons, making the axion–photon channel the leading decay mode of 
the dark photon.
This is the parameter space on which we focus in this paper. 

A heavy dark photon, without the existence of axion, has been searched at the CERN Large Hadron Collider, LHC, using $pp\to\gamma'$ with $\gamma'\to\mu^+\mu^-$ for a dark photon up to 70 GeV by the LHCb collaboration \cite{LHCb:2019vmc} and up to 200 GeV by the CMS collaboration \cite{CMS:2019buh}. The kinematic mixing parameter $\epsilon^2 \gtrsim 10^{-6}(10^{-5})$ is excluded for $m_{\gamma'}\simeq 10$ GeV ($70$ GeV)~\cite{LHCb:2019vmc}, and $\epsilon^2 \gtrsim 5\times10^{-5}$ as $m_{\gamma'}=200$ GeV~\cite{CMS:2019buh}. 
Searches for heavy dark photon at future lepton colliders have also been studied in Refs.~\cite{San:2022uud,Cheung:2025uaz}.

\section{ \label{sec:signature} Photons and Missing Energy at Lepton Colliders}

\begin{figure}[htbp!]
\centering
\includegraphics[width=0.85\textwidth]{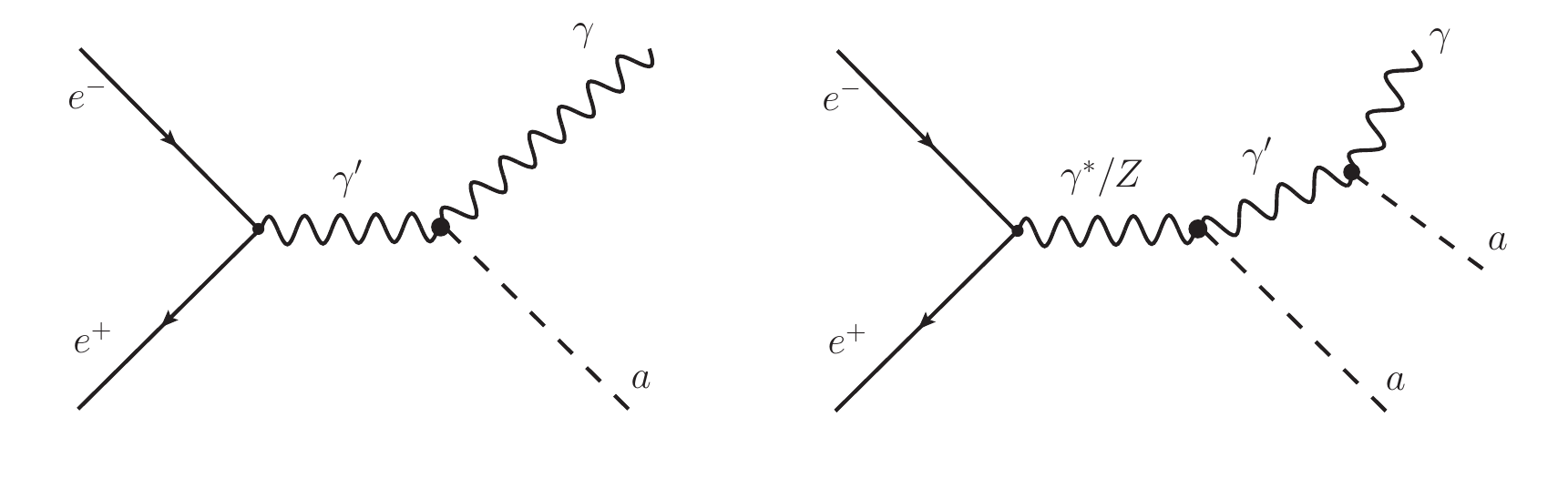}
\caption{\label{fig:feyndiags} Feynman diagrams for the single-photon plus missing energy signal.}
\end{figure}

The presence of axion--photon--dark photon interaction leads to photon signatures at $e^+ e^-$ colliders, particularly single-photon events with missing energy. 
These arise from processes such as $e^+ e^- \to \gamma' \to \gamma a$ and $e^+ e^- \to \gamma^*/Z \to a \gamma'$, followed by $\gamma' \to \gamma a$ as depicted in Fig.~\ref{fig:feyndiags}. If the axion is sufficiently light and weakly coupled to the SM particles, it escapes detection and causes missing energy. 
In this study, we first constrain the parameter space using LEP II data~\cite{L3:1999oof,L3:2003yon}.
Then we further consider future colliders, such as ILC, FCC-ee, and CEPC.

\begin{figure}[htbp!]
\centering
\includegraphics[width=0.5\textwidth]{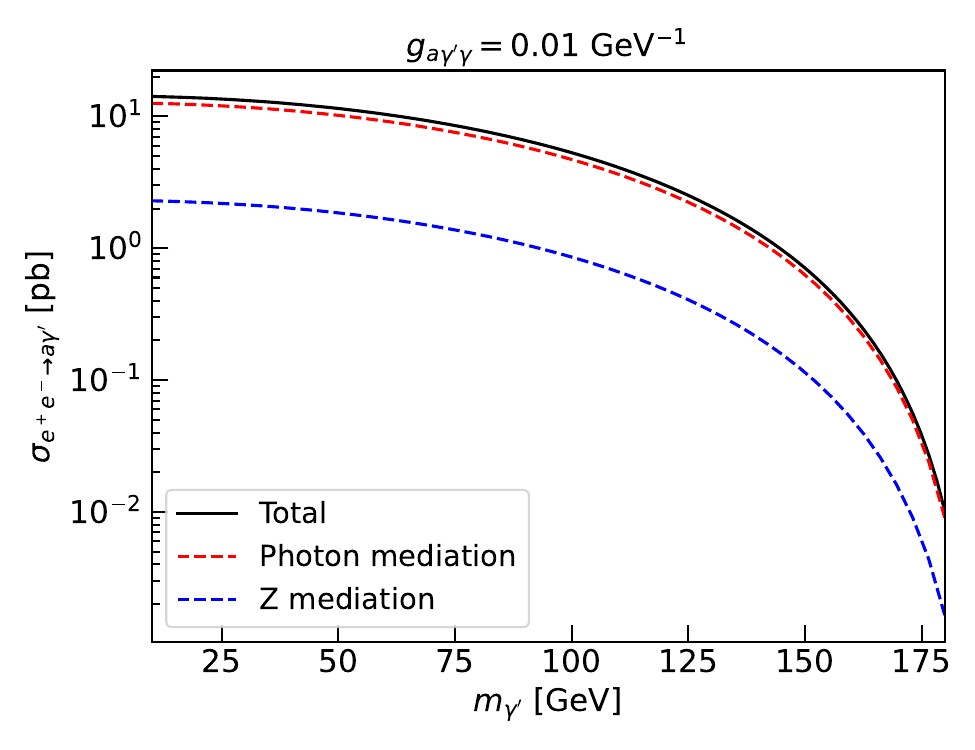}
\caption{\label{fig:LEP_xsec} 
The production cross section of $a\gamma'$ at LEP. The solid black line represents the total cross section, while the dashed red and blue lines indicate the individual contributions from photon and $Z$ boson exchange diagrams. We fix $g_{a\gamma'\gamma} = 0.01$ GeV$^{-1}$.} 
\end{figure}

\begin{figure}[htbp!]
\centering
\includegraphics[width=0.49\textwidth]{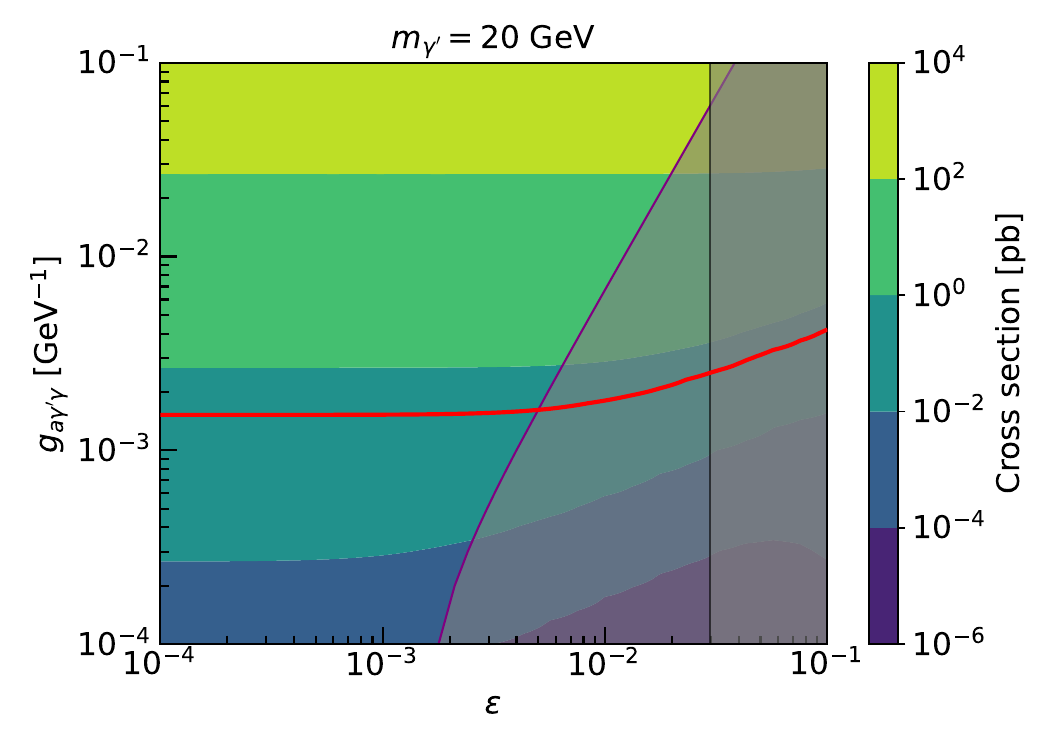}
\includegraphics[width=0.49\textwidth]{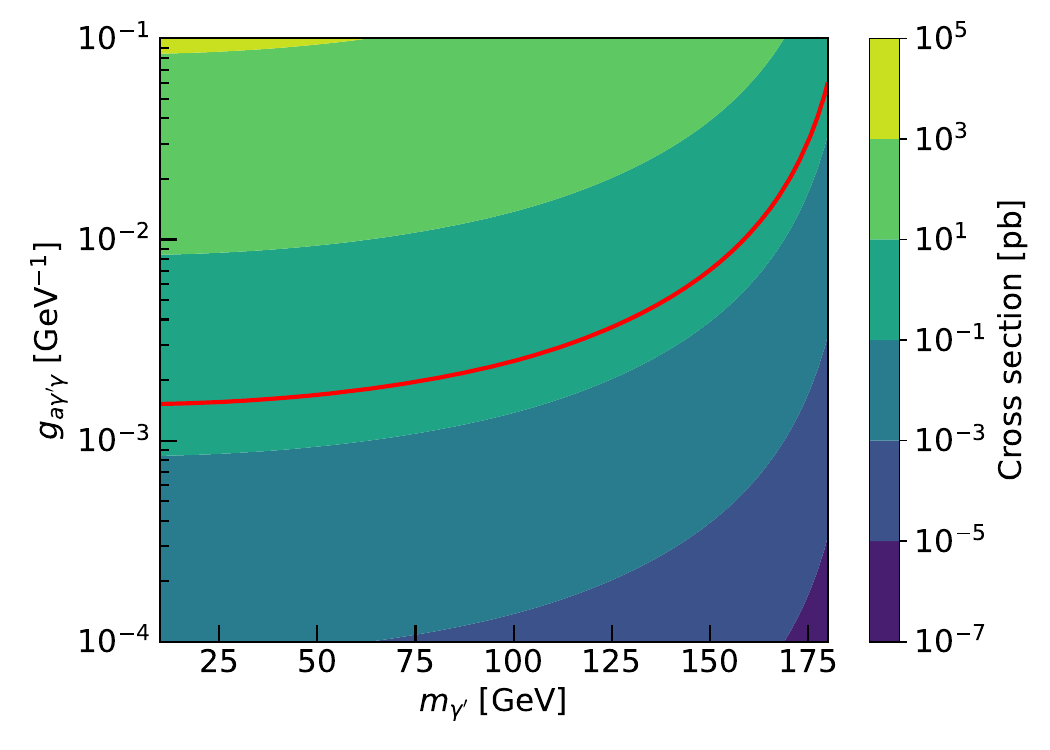}
\caption{\label{fig:LEPconstraint} 
 Constraints from LEP single photon data (red line) projected on the ($\epsilon$, $g_{a\gamma'\gamma}$) plane (left) and the ($m_{\gamma'}$, $g_{a\gamma'\gamma}$) plane (right). In the left panel, we fix $m_{\gamma'} = 20$ GeV while on the right panel, we set $\epsilon \ll m_{\gamma'}*g_{a\gamma'\gamma} $. The color gradient indicates the total cross section of the signal processes. The upper bound of cross section is shown with the red curve. The gray regions to the right of the solid purple and solid black curves are excluded by CMS dimuon searches~\cite{CMS:2019buh} and by precision measurements at the $Z$ pole~\cite{Curtin:2014cca, Du:2019mlc}, respectively.}
\end{figure}

\subsection{LEP II}
Single-photon events with missing energy have been studied at the LEP to search for supersymmetry and extra dimensions~\cite{L3:1999oof,L3:2003yon}\footnote{Multiple-photon plus missing energy events at LEP can also constrain the interactions in Eq.~(\ref{eq:Lag}), although these constraints are weaker than those from single-photon events (see Appendix~\ref{app:twophoton}).}. The results can be applied to the search for dark photons shown in Fig.~\ref{fig:feyndiags}. Two production channels contribute to signal events: (1) $e^+ e^- \to \gamma' \to \gamma a$ and (2) $e^+ e^- \to \gamma^*/Z \to a \gamma'$ with $\gamma' \to \gamma a$. 
It should be noted that channel (1) is proportional to $\epsilon^2$, while channel (2) is proportional to $g_{a\gamma'\gamma}^2$. Fig.~\ref{fig:LEP_xsec} shows the production cross section of $a\gamma'$ at LEP. The contribution from the $Z$-boson exchange diagram is significantly smaller than that from the off-shell photon exchange. This suppression arises from the weak mixing angle in the $Z a\gamma'$ coupling, as well as the additional propagator suppression from the $Z$-boson mass.    

We used the observed number of single-photon events at total energy $\sqrt{s} = 189$ GeV with selection of transverse momentum of photon $p_T(\gamma) > 0.02 \sqrt{s}$  from Table 2 of Ref.~\cite{L3:2003yon}. Assuming a detector efficiency of $80\%$ and a theoretical uncertainty of $1\%$, and applying the CLs method, we obtain a bound on the cross section for the single-photon plus the missing energy process of $\sim 0.32$ pb at the $95\%$ C.L. 
Applying this bound, we obtain constraints on $\epsilon$ and $g_{a\gamma'\gamma}$ for various $m_{\gamma'}$ assuming extremely light $m_a$. In our study, we take $m_a = 1$ MeV for simplicity. The left panel of Fig.~\ref{fig:LEPconstraint} shows the constraints of two couplings, $g_{a\gamma'
\gamma}$ and $\epsilon$ for $m_{\gamma'}=~20$ GeV. The region above the red curve is excluded since more single-photon events are generated than expected. 
We see that the cross section is insensitive to $\epsilon$ as $g_{a\gamma'\gamma}\gtrsim 10^{-3}~\rm{GeV}^{-1}$, which means that the process (2) dominates. 
 A slightly weaker bound is observed at large $\epsilon$ where the decay branching ratio Br$(\gamma' \to \gamma a)$ decreases.

Due to the presence of the kinetic mixing term, characterized by the parameter $\epsilon$ in Eq.~(\ref{eq:Lag}), the $Z$ boson mass and its couplings are modified. 
These deviations are tightly constrained by precision $Z$-pole measurements at LEP~\cite{Curtin:2014cca, Du:2019mlc}. 
In addition, searches for narrow resonances decay into dileptons at the LHC impose 
bounds on dark photon production and its branching ratio into leptons.  
We recast the constraints from the CMS dimuon search~\cite{CMS:2019buh} 
accordingly (see Appendix~\ref{app:CMSrecast}).
The resulting exclusions projected in $(\epsilon,\,g_{a\gamma'\gamma})$ plane are shown as the gray regions in the left panel 
of Fig.~\ref{fig:LEPconstraint} for a representative dark photon mass 
$m_{\gamma'} = 20~\text{GeV}$.
Combining CMS dilepton searches, $Z$-pole, and LEP data, we find the viable parameter space is $g_{a\gamma'\gamma} \lesssim 1.5 \times 10^{-3}~\mathrm{GeV}^{-1}$ and $\epsilon \lesssim 2.0 \times 10^{-3}$.

The right panel of Fig.~\ref{fig:LEPconstraint} shows the projection of the LEP constraints on the $(m_{\gamma'}, g_{a\gamma'\gamma})$ plane, assuming that $\epsilon \ll m_{\gamma'} g_{a\gamma' \gamma}$ satisfied the $Z$-pole and CMS dimuon searches constraints. Also, the red curve shows the upper limit of the single-photon plus missing energy production cross section\footnote{We note that this constraint from LEP excludes the region that explains the muon $g-2$ anomaly studied in \cite{Chen:2024jbr}.
}.

\begin{figure}[htbp!]
\centering
\includegraphics[width=0.49\textwidth]{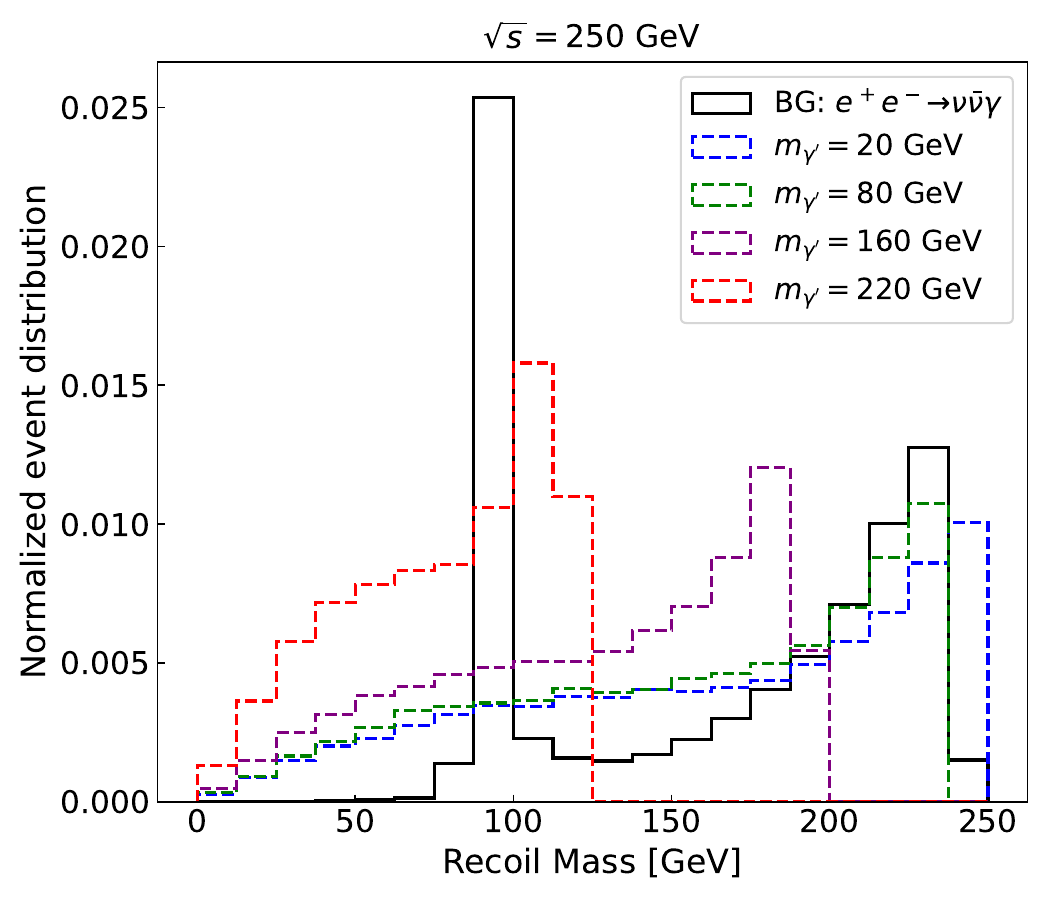}
\includegraphics[width=0.49\textwidth]{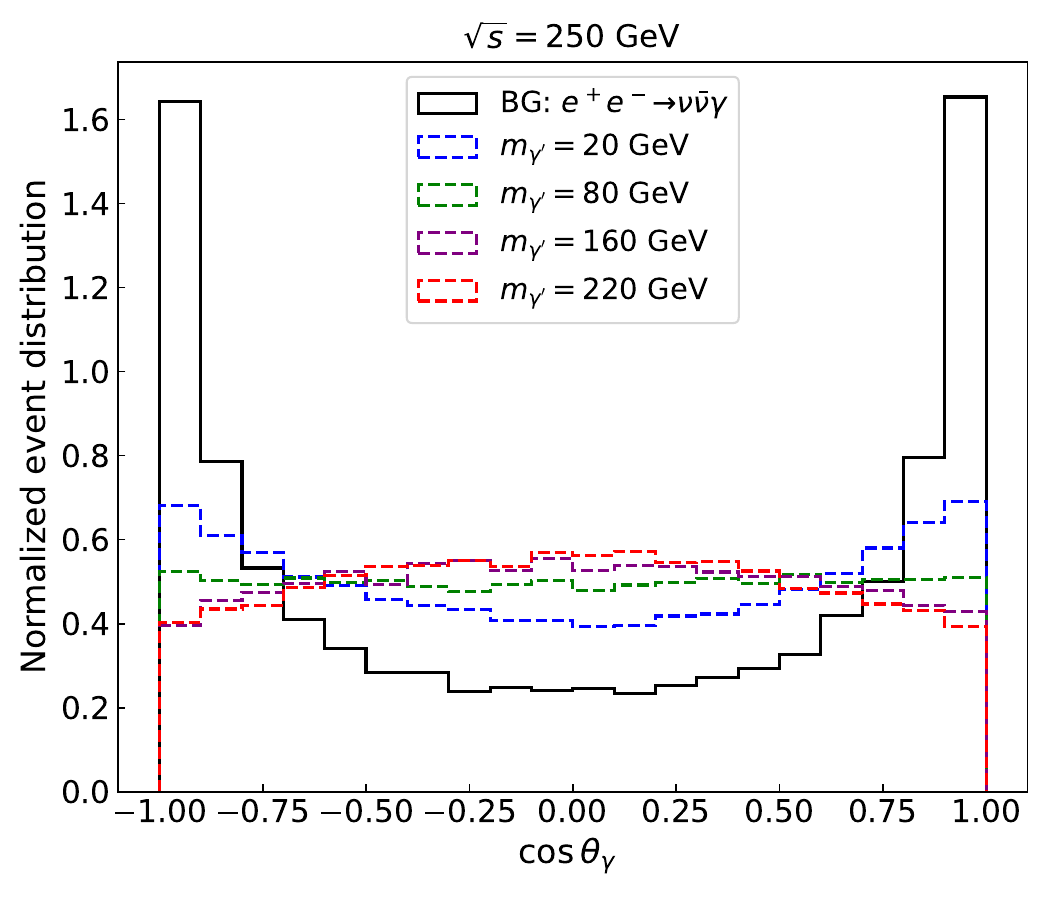}
\caption{\label{fig:distribution} Recoil mass distribution (left) and photon polar angle distribution (right). The solid black histogram represents the SM background; dashed curves correspond to signal with $m_{\gamma'} = 20, 80, 160, 200$ GeV (blue, green, purple, red).}
\end{figure}

\begin{table}[h!]
\centering
\begin{tabular}{|l|c|c|l|}
\hline
\textbf{Collider} & $\sqrt{s}$ (GeV) & Luminosity (ab$^{-1}$) & Notes \\
\hline
ILC~\cite{LCCPhysicsWorkingGroup:2019fvj}     & 250 & 2.0 & 10-year Higgs factory phase \\
FCC-ee~\cite{FCC:2018evy}  & 240 & 5.0 & 3-year Higgs factory phase \\
CEPC~\cite{CEPCStudyGroup:2023quu}    & 240 & 13.0 & 10-year operation with 30 MW SR power \\
\hline
\end{tabular}
\caption{Projected luminosities at future $e^+e^-$ colliders.}
\label{tab:collider-luminosity}
\end{table}

\subsection{Future $e^+ e^-$ Colliders}

We analyze the single-photon signal with missing energy at future $e^+ e^-$ colliders. Simulations of signal and SM background processes are performed using \texttt{MadGraph5}~\cite{Alwall:2014hca}. The dominant background is $e^+ e^- \to \nu\bar{\nu}\gamma$, with a leading-order cross section of $\sim 1.56$ pb at $\sqrt{s} = 250$ GeV.

The distributions of recoil mass and photon polar angle are shown in Fig.~\ref{fig:distribution}. A characteristic drop-off is seen in the  histogram of recoil mass, which is related to the mass of the dark photon. For heavier dark photons, the photon from dark photon decay is more energetic. Therefore, the upper bound of the recoil mass is smaller, according to the definition $M_\text{recoil}^2 = s - 2 \sqrt{s} E_\gamma$, where $E_\gamma$ is the energy of the observed photon. We also notice that the recoil mass distribution exhibits sharp edges at $M_\text{recoil}^{\text{max}}$, allowing the dark photon mass to be reconstructed via:
\begin{equation}
m_{\gamma'} = \sqrt{s - \left(M_\text{recoil}^{\text{max}}\right)^2}.
\end{equation}
Two peaks are shown in the SM background: the peak at the $Z$-boson mass comes from $e^+e^-\to \gamma Z(\to \nu\bar{\nu})$, while another one close to the total energy is contributed by a three-body final state of $e^+e^-\to \gamma\nu_e\bar{\nu}_e$ through a $W$-boson mediated in the t-channel and with initial-state radiations.  
Heavier dark photons also lead to broader angular distributions of the emitted photon. In contrast, SM background photons are concentrated near the beamline.

\begin{figure}[htbp!]
\centering
\includegraphics[width=0.55\textwidth]{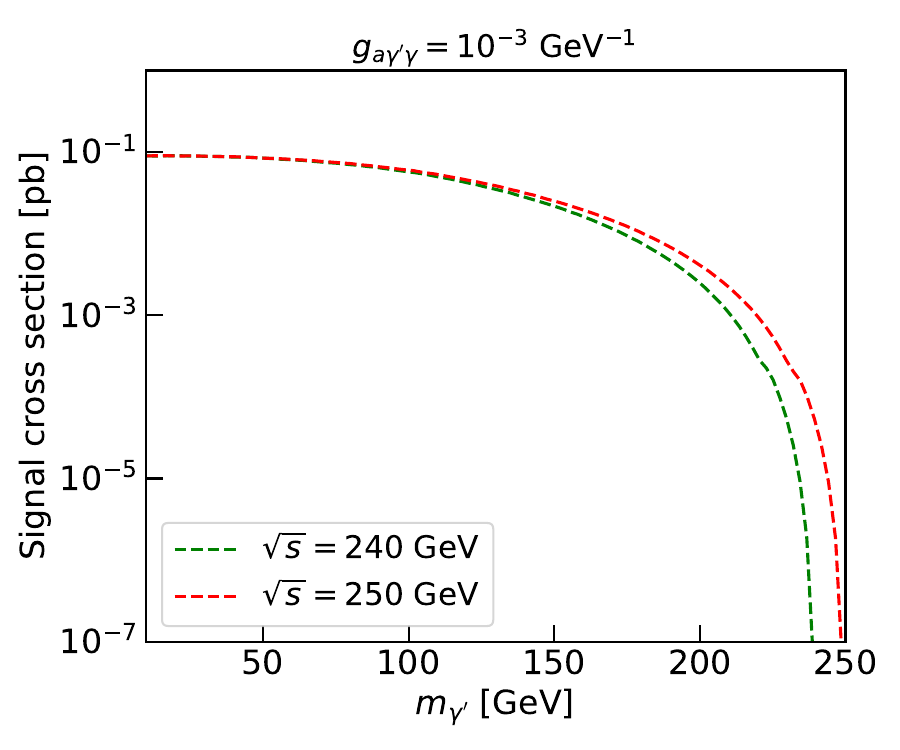}
\caption{\label{fig:xsec} Signal cross section after selection cuts for $\sqrt{s} = 240$ GeV (green dashed) and 250 GeV (red dashed). We fix $g_{a\gamma'\gamma} = 10^{-3} \mathrm{GeV}^{-1}$, $m_a = 1$ MeV, and $\epsilon \ll g_{a\gamma'\gamma}m_{\gamma'}$.}
\end{figure}

\begin{figure}[htbp!]
\centering
\includegraphics[width=0.55\textwidth]{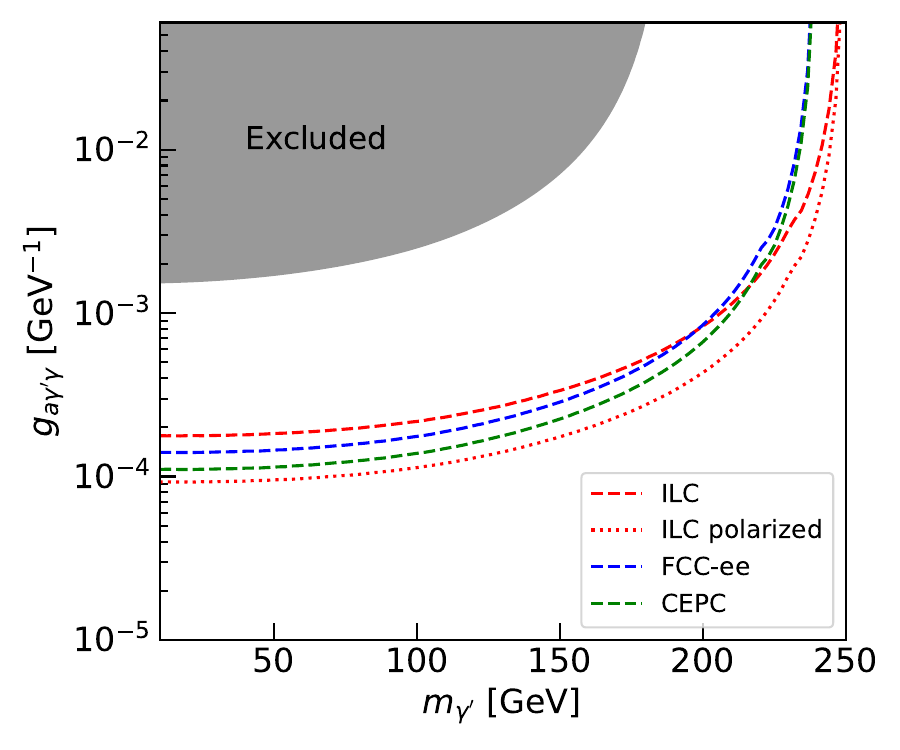}
\caption{\label{fig:sensitivity} Sensitivity projections at ILC (red dashed), FCC-ee (blue dashed), and CEPC (green dashed) on the ($m_{\gamma'}$, $g_{a\gamma'\gamma}$) plane. The red dotted line denotes the projected sensitivity at the ILC with longitudinally polarized beams (see Sec.~\ref{sec:pol} for details). The gray region is excluded by LEP II data.}
\end{figure}

To optimize the signal-to-background ratio, we apply the following cuts:
\begin{align}
M_{\rm recoil} &\notin [ m_Z - 2 \Gamma_Z, m_Z + 2 \Gamma_Z ], \label{eq:cut1} \\
|\cos\theta_\gamma| &< 0.75, \label{eq:cut2}
\end{align}
where $m_Z$ and $\Gamma_Z$ are the $Z$-boson mass and the total decay width, respectively. 
After applying these cuts, the background is reduced by $\sim65\%$, while the pass efficiency of the dark photon signal ranges from $65\%$ to $80\%$ depending on $m_{\gamma'}$. The resulting cross sections after selections for $m_{\gamma'}
\ge 20$ GeV are shown in Fig.~\ref{fig:xsec}. 
For $m_{\gamma'}\lesssim 100$ GeV, the signal cross sections remain almost unchanged, since the total energy $\sqrt{s}$ is much higher than $m_{\gamma'}$.
We compute event yields as $N_{s,b} = \sigma_{s,b} \times \mathcal{L}$ is the number of events after the selections with the lower-scripts {\it s} and {\it b} representing the dark-photon signals and the SM background, respectively, and ${\cal L}$ is the integrated luminosity. Taking the projected luminosity shown in Tab.~\ref{tab:collider-luminosity} for different colliders in the future,
we then evaluate the significance $\mathcal{Z} = N_s/\sqrt{N_b}$. Requiring $\mathcal{Z} = 5$, we obtain projected sensitivities for ILC, FCC-ee, and CEPC shown in Fig.~~\ref{fig:sensitivity}. These machines are sensitive to the coupling $g_{a\gamma'\gamma}$ down to $\sim 10^{-4},\mathrm{GeV}^{-1}$ for $m_{\gamma'} \sim \mathcal{O}(10\text{--}100~~\mathrm{GeV})$. For heavier dark photons with $m_{\gamma'}\gtrsim 200$ GeV, these $e^+e^-$ colliders quickly fail to probe $g_{a\gamma'\gamma}$ with single-photon events due to the limited total energy.

\section{Polarization search at the ILC}
\label{sec:pol}
Beam polarization is one of the key advantages of the ILC which can enhance the sensitivity to the 
$e^+e^- \to a\,\gamma'$ signal.  
Let $P_{e^-}$ and $P_{e^+}$ denote the longitudinal polarizations of the 
electron and positron beams.  The polarized cross section is given by~\cite{Moortgat-Pick:2005jsx}
\begin{align}
\sigma(P_{e^-},P_{e^+}) = \frac{1}{4} \big[
&(1+P_{e^-})(1+P_{e^+})\,\sigma_{RR}
+(1-P_{e^-})(1-P_{e^+})\,\sigma_{LL} \nonumber \\
&+(1+P_{e^-})(1-P_{e^+})\,\sigma_{RL}
+(1-P_{e^-})(1+P_{e^+})\,\sigma_{LR}
\big],
\end{align}
where $\sigma_{ij}$ denotes the cross section for initial helicities 
$i$ (electron) and $j$ (positron).

For an $s$-channel process mediated by a spin--1 gauge boson, the total 
angular momentum along the beam axis must satisfy $J_z=\pm 1$.  
This implies that only opposite-helicity configurations contribute, 
while the same-helicity ($LL$ and $RR$) states vanish in the 
ultrarelativistic limit~\cite{Moortgat-Pick:2005jsx}.

For the SM background $e^+e^-\to\nu\bar{\nu}\gamma$, the dominant 
contribution arises from $t$-channel $W$ exchange, which couples only to 
left-handed electrons and right-handed positrons. Whereas, the 
$s$-channel $Z$ exchange contributes almost equally to both the $LR$ and $RL$ channels.
Thus, the $LR$ configuration provides the largest contribution.
The same-helicity cross sections vanish in both SM background channels. 
At leading order and $\sqrt{s}=250$~GeV, we find
$\sigma^{\rm BG}_{LR} \simeq 8.17~{\rm pb}$ and 
$\sigma^{\rm BG}_{RL} \simeq 2.70~{\rm pb}$.

\begin{figure}[htbp!]
\centering
\includegraphics[width=0.55\textwidth]{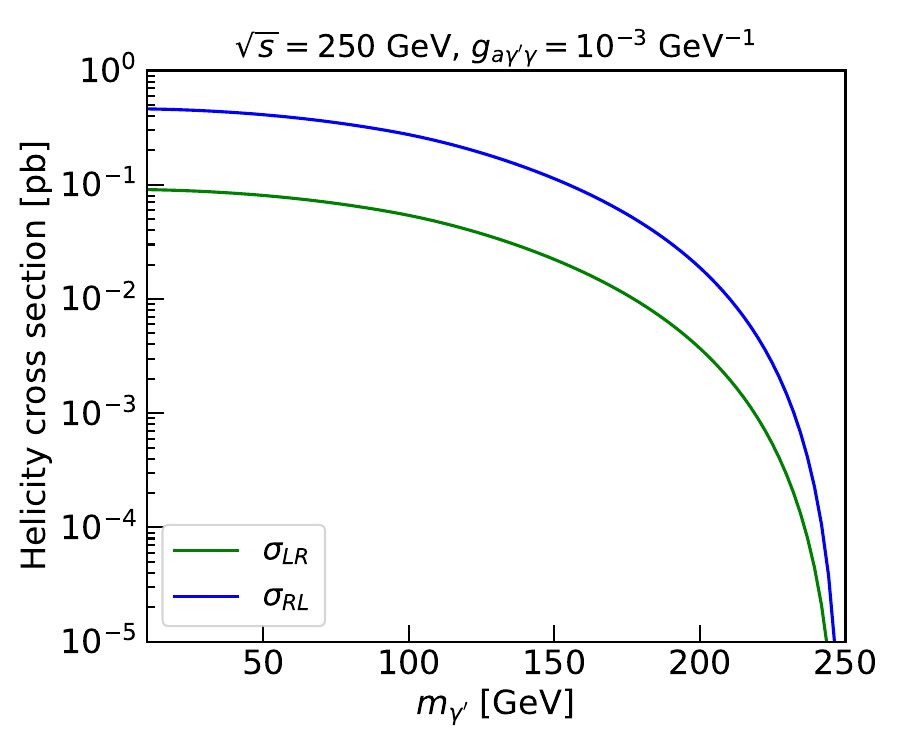}
\caption{\label{fig:xsec_pol} Helicity cross sections for the  signal processes at $\sqrt{s} = 250$ GeV. The green line represents the cross section for initial left-handed electron and right-handed positron ($\sigma_{LR}$), and the blue line indicates the cross section for initial right-handed electron and left-handed positron ($\sigma_{RL}$).} 
\end{figure}

In contrast, the signal exhibits the opposite helicity preference.  
As shown in Fig.~\ref{fig:xsec_pol}, we find 
$\sigma^{\rm sig}_{RL} \simeq 5 \times \sigma^{\rm sig}_{LR}$ for the dominant 
signal process $e^+e^- \to a\,\gamma'$ (with $\gamma' \to a\gamma$) at 
$\sqrt{s}=250$~GeV.  
The difference between the $LR$ and $RL$ signal cross sections originates from 
the interference between the photon- and $Z$-exchange amplitudes.  
For $\sqrt{s} > m_Z$, this interference is constructive in the $RL$ channel and 
destructive in the $LR$ channel, leading to an enhancement of 
$\sigma^{\rm sig}_{RL}$ and a suppression of $\sigma^{\rm sig}_{LR}$.

Since the background is dominated by the $LR$ channel, while the signal 
is largest in the $RL$ channel, the optimal strategy is to enhance the 
$RL$ contribution and suppress the $LR$ one.  
Within the standard ILC baseline capabilities~\cite{ILC:2013jhg}, a 
representative choice is
$P_{e^-} \simeq +0.8$ and $P_{e^+} \simeq -0.3$
which increases the weight of $\sigma_{RL}$ and simultaneously reduces 
that of $\sigma_{LR}$.

\begin{figure}[htbp!]
\centering
\includegraphics[width=0.49\textwidth]{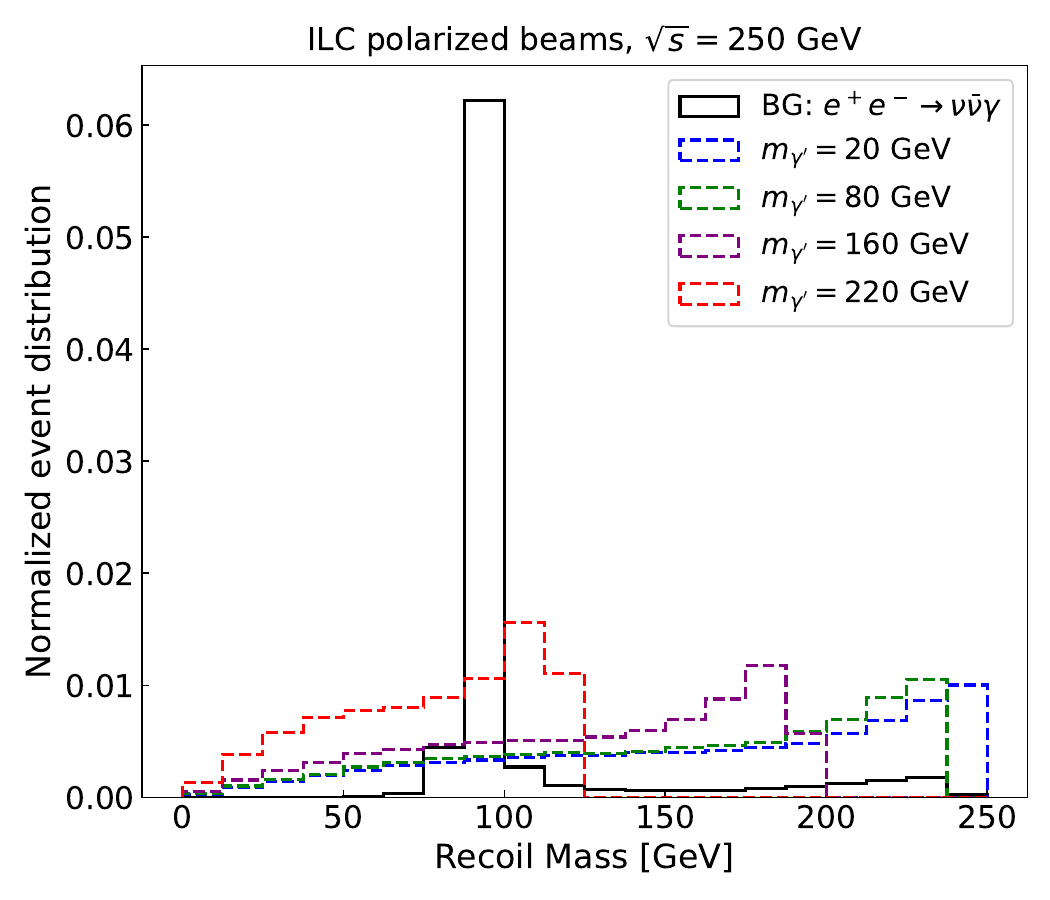}
\includegraphics[width=0.49\textwidth]{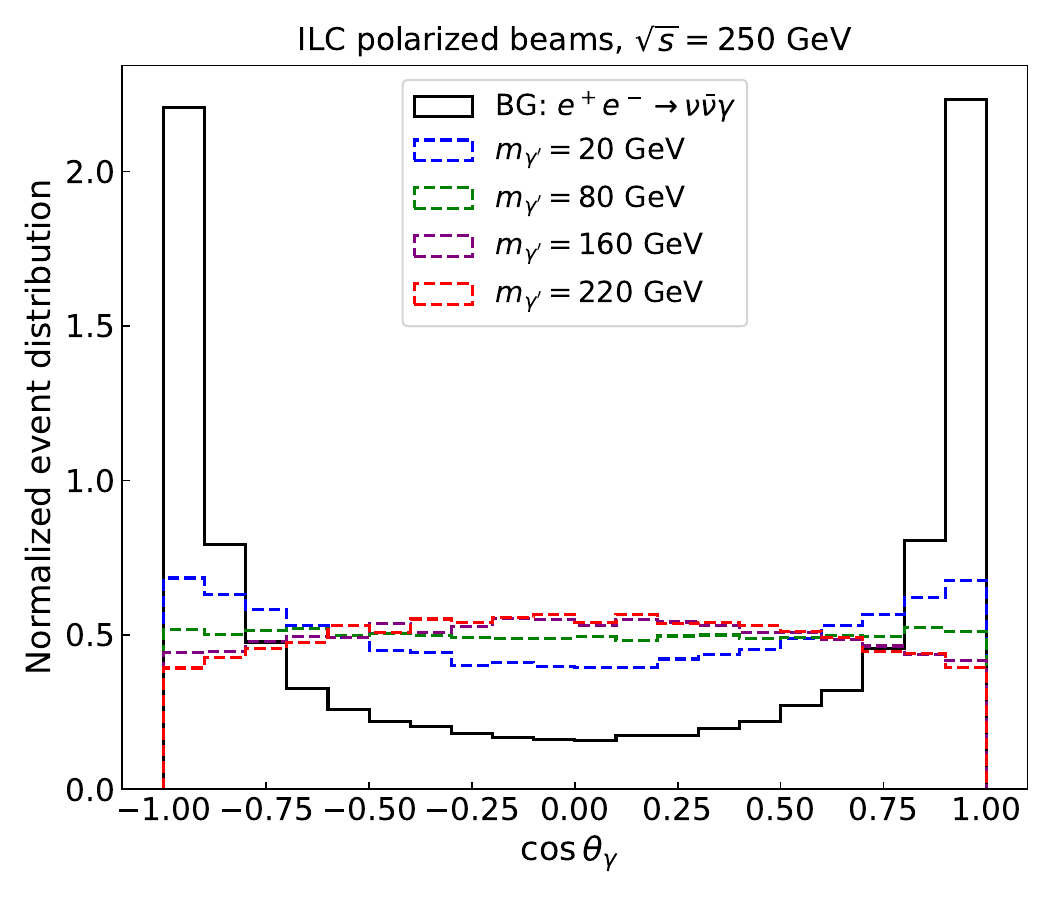}
\caption{\label{fig:distribution_pol} Same as Fig.~\ref{fig:distribution}, but with longitudinal polarized beams 
$P_{e^-}=+0.8$ and $P_{e^+}=-0.3$ at the ILC.
}
\end{figure}

In Fig.~\ref{fig:distribution_pol}, we show the normalized recoil-mass and photon polar angle distributions for the ILC 
beam-polarized setup.  
Compared to the unpolarized case in Fig.~\ref{fig:distribution}, the 
high recoil mass background events originating from $t$-channel $W$ 
exchange are significantly suppressed, while the signal distribution 
remains largely unchanged.  
The photon polar angle distribution is also very similar to the 
unpolarized case.

\begin{figure}[htbp!]
\centering
\includegraphics[width=0.55\textwidth]{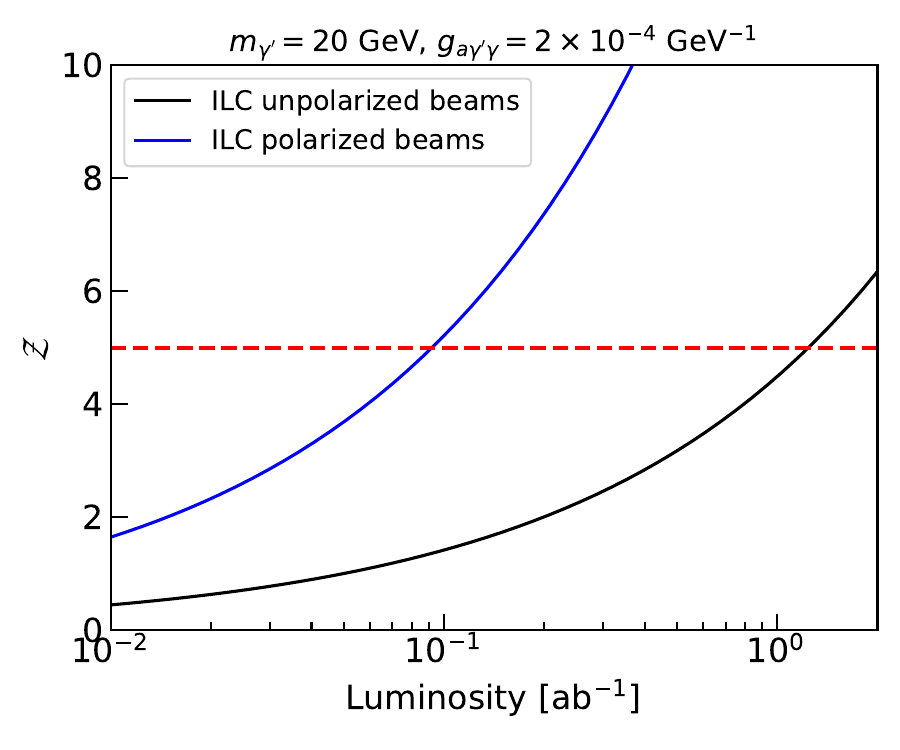}
\caption{\label{fig:lumi} Significance ${\cal Z}$ as a function of the integrated luminosity at the ILC with fixing $m_{\gamma'} = 20$ GeV and $g_{a\gamma'\gamma} = 2\times 10^{-4}$ GeV$^{-1}$. 
The solid black and blue curves correspond to the single-photon plus 
missing-energy search with unpolarized and polarized beams, respectively. 
The dashed red line indicates the ${\cal Z}=5$ threshold. 
} 
\end{figure}

Using the same kinematic cuts in 
Eqs.~(\ref{eq:cut1}) and (\ref{eq:cut2}), we evaluate the improvement in 
sensitivity from beam polarization.  
We find that the signal significance increases by a factor of about four 
compared to the unpolarized search.  
Fig.~\ref{fig:lumi} shows the significance ${\cal Z}$ as a function of 
the integrated luminosity at the ILC for a benchmark point 
$m_{\gamma'} = 20~\text{GeV}$ and 
$g_{a\gamma'\gamma} = 2\times 10^{-4}~\text{GeV}^{-1}$.  
For this parameter space choice, achieving ${\cal Z}=5$ requires only 
${\cal L} \simeq 0.1~\text{ab}^{-1}$ with polarized beams, 
whereas more than an order of magnitude larger luminosity is needed for 
the unpolarized case.

The resulting sensitivity is shown as the dotted red curve in 
Fig.~\ref{fig:sensitivity} on the 
$(m_{\gamma'},\,g_{a\gamma'\gamma})$ plane, providing the strongest reach 
into the parameter space.

\section{\label{sec:conclusion}  Conclusion}

We have explored the potential of future lepton colliders to probe the interaction between axions, dark photons, and Standard Model photons through dimension-five operators. Such interactions can naturally arise in models with light pseudoscalar particles and dark photons, and may be generated at the loop level via heavy fermions charged under both $U(1)_{\rm Y}$ and $U(1)_D$.

Focusing on single-photon events with missing energy, we analyzed the signal processes $e^+ e^- \to \gamma' \to \gamma a$ and $e^+ e^- \to \gamma^*/Z \to a \gamma' \to a a \gamma$, under the assumption of a light axion that escapes detection. The axion--photon--dark photon coupling $g_{a\gamma'\gamma}$ larger than about $10^{-3}~\rm{GeV}^{-1}$ is excluded by LEP II data for $m_{\gamma'}\lesssim 100$ GeV. We studied the relevant kinematic distributions and developed selection strategies to optimize the signal-to-background ratio. In particular, the recoil mass and the angular distribution of the emitted photon were found to be especially effective in discriminating signals from background.

Using these strategies, we projected the sensitivity reach of ILC, CEPC, and FCC-ee. We found that these future facilities can probe the coupling $g_{a\gamma'\gamma}$ in the range of $\sim 10^{-4}~\mathrm{GeV}^{-1}$ to $10^{-3}~\mathrm{GeV}^{-1}$ for dark photon masses between 10 GeV and 230 GeV. 
Furthermore, we demonstrated that the photon energy spectrum offers a viable method to reconstruct the dark photon mass through recoil kinematics, providing an additional handle on signal characterization.

We also analyzed the signal using beam 
polarization at the ILC.  
We found that the signal and background exhibit markedly different 
helicity structures: the SM background is dominated by the $LR$ (left-handed electron and right-handed positron) initial 
state due to the dominant contributions of $t$-channel $W$ exchange, 
while the signal is enhanced in the $RL$ (right-handed electron and left-handed positron) initial state due to 
constructive photon--$Z$ interference.  
Exploiting the standard ILC polarization configuration,
we found that the signal significance improves by approximately a factor 
of four relative to the unpolarized case, with the discovery reach 
achievable using only ${\cal O}(0.1)\,\text{ab}^{-1}$ of data for a benchmark point 
$m_{\gamma'} = 20~\text{GeV}$ and 
$g_{a\gamma'\gamma} = 2\times 10^{-4}~\text{GeV}^{-1}$.
The resulting polarized sensitivity 
provides the best projected reach in the 
$(m_{\gamma'}, g_{a\gamma'\gamma})$ parameter space among all 
$e^+e^-$ collider options considered. 

Our results highlight the importance of high-luminosity lepton colliders in exploring light and weakly coupled new physics, offering complementary coverage to fixed-target, astrophysical, and hadronic collider searches. This work paves the way for probing rich dynamics within the dark sector at next-generation facilities.

Finally, we note that the existing LHC monophoton searches can provide a complementary constraint on the axion–dark-photon interaction, particularly for heavier dark photon regions.
A dedicated detector-level analysis, as well as projections for the HL-LHC and HE-LHC, could significantly extend the sensitivity into the higher-mass region of the dark photon parameter space and is left for future work.

\section*{Acknowledgments}
This work was partially supported by the National Science and Technology Council (NSTC) of Taiwan under Grant No. NSTC-113-2112-M-003-007 and NSTC-114-2112-M-003-009 (CRC), the Ministry of Education (Higher Education Sprout Project NTU-114L104022-1), and the National Center for Theoretical Sciences of Taiwan (VQT).
This work was also supported in part by the Vietnam National Foundation for Science and Technology Development (NAFOSTED) under grant number 103.01-2023.50 (VQT).

\appendix

\begin{figure}[htbp!]
\centering
\includegraphics[width=0.85\textwidth]{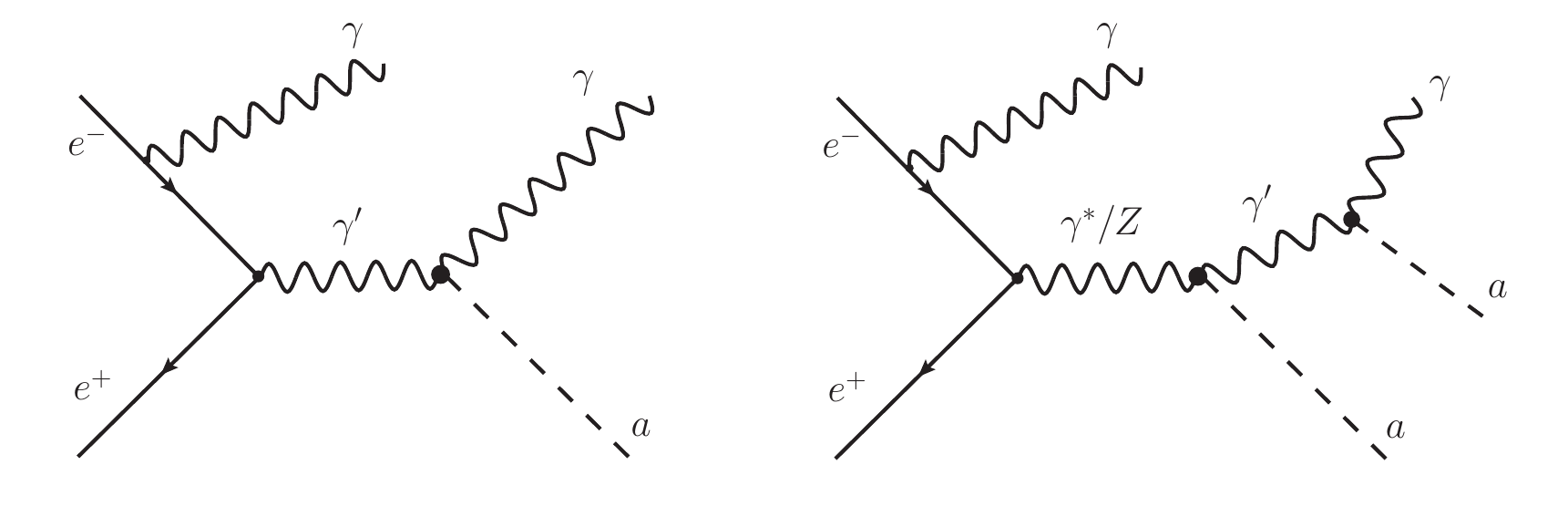}
\caption{\label{fig:feyndiags2} Feynman diagrams for the two photon plus missing energy signal.}
\end{figure}

\begin{figure}[htbp!]
\centering
\includegraphics[width=0.65\textwidth]{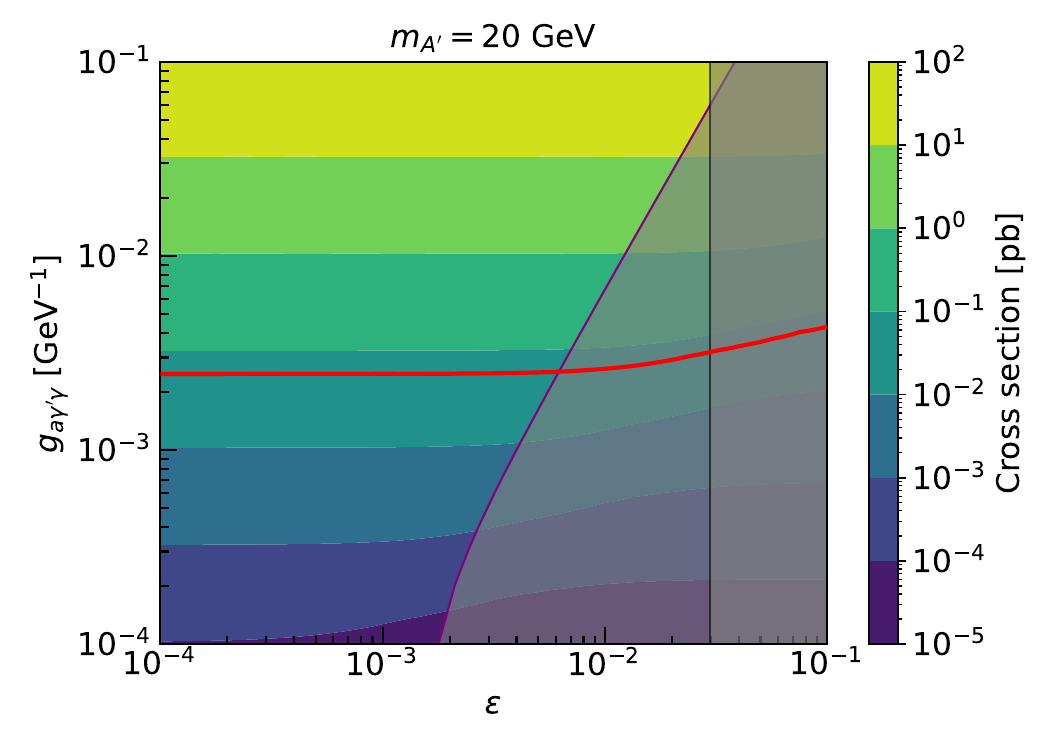}
\caption{\label{fig:LEPconstraint_2photon} 
Constraints from LEP data of two photon plus missing energy (red line) projected on the ($\epsilon$, $g_{a\gamma'\gamma}$) plane. We fix $m_{\gamma'} = 20$ GeV. The color gradient indicates the total cross section of the two-photon signal processes. 
The gray regions to the right of the solid purple and solid black curves are excluded by CMS dimuon searches~\cite{CMS:2019buh} and by precision measurements at the $Z$ pole~\cite{Curtin:2014cca, Du:2019mlc}, respectively.
}
\end{figure}

\section{LEP Constraints from Two-Photon plus Missing Energy}
\label{app:twophoton}
Searches for multi-photon final states with missing energy at LEP can also be 
used to constrain the interaction in Eq.~(\ref{eq:Lag}).  
In our model, the relevant signal topologies are shown in 
Fig.~\ref{fig:feyndiags2}.  
We assume that the axion is sufficiently light and couples only weakly to the SM 
sector, so that it escapes the detector and contributes to the missing-energy 
signature. 

In the parameter space considered in this analysis, the contribution from the
signal process shown in the left panel of Fig.~\ref{fig:feyndiags2} is
subdominant.
This is because dark photon production in this channel is proportional to
$\epsilon^2$.
Enhancing this contribution would require a large kinetic mixing parameter
$\epsilon$, which simultaneously suppresses the branching ratio of the dark
photon decay into the $a\gamma$ final state.
Moreover, such large values of $\epsilon$ are already excluded by CMS dilepton
searches and precision $Z$-pole measurements.

In contrast, the production of dark photons in the topology shown in the right
panel of Fig.~\ref{fig:feyndiags2} is proportional to
$g_{a\gamma'\gamma}^2$.
A larger value of this coupling not only enhances the production rate but also
increases the branching fraction of the dark photon decay into the
photon–axion final state, making this channel more relevant for LEP searches.

To recast the LEP bounds, we use the L3 data collected at a center-of-mass 
energy $\sqrt{s} = 189$~GeV, with the event selection requiring at least two photons with energy above 1 GeV and a
global transverse momentum of photons $P_T > 0.02\sqrt{s}$~\cite{L3:2003yon}.  
We assume a detector efficiency of 80\% and a theoretical uncertainty of 1\%.  
Using the CL$_s$ method, we obtain an upper limit on the total cross section 
for the two photon plus missing energy process of 
$0.057~\text{pb}$ at 95\%~C.L.

This bound is translated into a constraint on 
$(\epsilon,\,g_{a\gamma'\gamma})$ for $m_{\gamma'} = 20~\text{GeV}$, as shown 
in Fig.~\ref{fig:LEPconstraint_2photon}.
We find that the resulting constraint from the two photon channel is weaker by 
roughly a factor of two compared to the single photon limit.

\section{Recasting the LHC Dilepton Limit}
\label{app:CMSrecast}
In our model, the branching fractions of the dark photon differ from those in the conventional dark photon scenario due to the presence of new decay channels 
such as $\gamma' \to a\gamma$ and $\gamma' \to a Z$, whenever kinematically 
allowed.  
Therefore, when recasting the LHC constraints from narrow-resonance dilepton searches, these modified branching ratios must be taken into account~\cite{Chen:2024jbr}. 
In particular, the bound on the kinetic mixing parameter must be rescaled to 
account for the reduced branching ratio of $\gamma'$ decays into leptons.  
The modified bound on the kinetic mixing $\epsilon_{\rm bound}$ can be obtain by 
\begin{equation}
\label{eq:witheps}
\epsilon_{\rm bound}^2 \, {\rm BR} (\gamma' \to \ell^+\ell^-) 
=  \epsilon_0^2 \, {\rm BR}_0(\gamma' \to \ell^+\ell^-),
\end{equation}
where $\epsilon_0$ is the LHC bound on the kinetic mixing in the 
conventional dark photon model~\cite{CMS:2019buh}, and 
${\rm BR}_0(\gamma' \to \ell^+\ell^-)$ is the corresponding branching ratio 
without additional decay channels.

Applying this procedure, we obtain the constraint on $\epsilon$ shown by the 
solid purple curves in Figs.~(\ref{fig:LEPconstraint}) and 
(\ref{fig:LEPconstraint_2photon}) for $m_{\gamma'} = 20~\text{GeV}$ using the dimuon searches data at CMS~\cite{CMS:2019buh}.  
As the coupling $g_{a\gamma'\gamma}$ increases, the dimuon branching fraction 
of the dark photon is progressively suppressed, leading to a correspondingly 
weaker bound on the kinetic mixing parameter.

\allowdisplaybreaks
\bibliographystyle{apsrev4-1}
\bibliography{refs}

\end{document}